# Data Navigator: An Accessibility-Centered Data Navigation Toolkit

Frank Elavsky, Lucas Nadolskis, Dominik Moritz

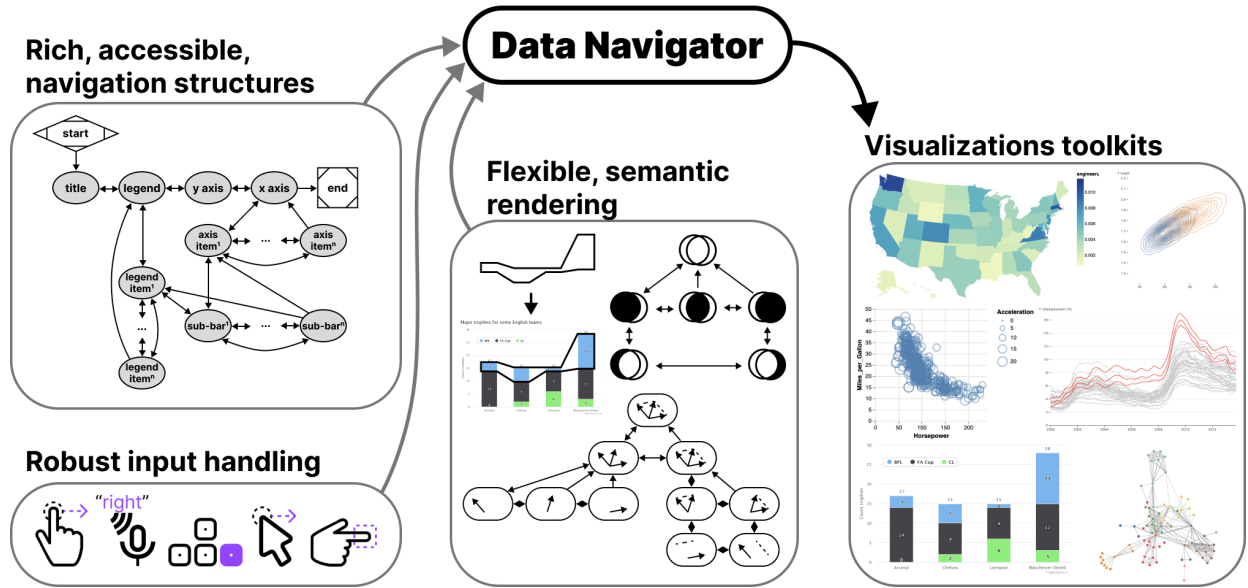

Fig. 1: Data Navigator provides data visualization libraries and toolkits with accessible data navigation structures, robust input handling, and flexible semantic rendering capabilities.

**Abstract**—Making data visualizations accessible for people with disabilities remains a significant challenge in current practitioner efforts. Existing visualizations often lack an underlying navigable structure, fail to engage necessary input modalities, and rely heavily on visual-only rendering practices. These limitations exclude people with disabilities, especially users of assistive technologies. To address these challenges, we present Data Navigator: a system built on a dynamic graph structure, enabling developers to construct navigable lists, trees, graphs, and flows as well as spatial, diagrammatic, and geographic relations. Data Navigator supports a wide range of input modalities: screen reader, keyboard, speech, gesture detection, and even fabricated assistive devices. We present 3 case examples with Data Navigator, demonstrating we can provide accessible navigation structures on top of raster images, integrate with existing toolkits at scale, and rapidly develop novel prototypes. Data Navigator is a step towards making accessible data visualizations easier to design and implement.

---

## 1 INTRODUCTION

While there is a growing interest in making data visualizations more accessible for people with disabilities, current toolkit and practitioner efforts have not risen to the challenge at scale. Major data visualization tools and ecosystems predominantly produce inaccessible artifacts for many users with disabilities. We believe this is largely a gap caused by a lack of underlying structure in most visualizations, failure to engage the input modalities used by people with disabilities, and over-reliance on visual-only rendering practices.

Users who are blind or low vision commonly use screen readers and users with motor and dexterity disabilities often do not use "pointer" (precise mouse and touch) based input technology when interacting with digital interfaces. Many users with motor and dexterity disabilities use discrete navigation controls, either sequentially using keyboard-like input, or directly using voice or text commands.

Most interactive visualizations simply focus on pointer-based input: they can be clicked or tapped, hovered, and selected in order to perform analytical tasks. This excludes non-pointer input technologies. These devices require consideration for the navigation structure and underlying semantics of a visual interface.

However, building navigable spatial and relational interfaces is a difficult task with current resources.

Raster images, arguably the most common format for creating and disseminating data visualizations, currently cannot be made into navigable structures. These are only described using alt text, which limits their usefulness to screen reader users.

Unfortunately, more accessible rendering formats like SVG with ARIA (accessible rich internet applications) properties are more resource intensive than raster approaches, like WebGL-powered HTML canvas or pre-rendered PNG files. SVG puts a burden on low-bandwidth users and a ceiling on how many data points can be rendered in memory.

In addition, ARIA itself has 2 major limitations. First, when added to interface elements, ARIA only provides *screen reader* access, which means that developers must build a solution from scratch for other navigation input modalities. Second, ARIA's linear navigation structure can be time-consuming for screen reader users if a visualization has many elements. This may impede how essential insights and relationships

---

- *All authors are with Carnegie Mellon University.*
- *Frank Elavsky: fje@cmu.edu.*
- *Lucas Nadolskis: nadolskis@cmu.edu.*
- *Dominik Moritz: domoritz@cmu.edu.*

are understood [14, 19, 32, 37, 38, 47].

Some emerging approaches have sought to address this serial limitation of data navigation and provide richer experiences for screen reader users [14, 37, 38, 47]. However, these approaches rely on a tree-based navigation structure which is often not an appropriate choice for visualizations of relational, spatial, diagrammatic, or geographic data. Many visualization structures are currently unaddressed.

Zong et al. stress that in order to realize richer, more accessible data visualizations, the responsibility must be shared by "toolkit makers," the practitioners who design, build, and maintain visualization authoring technologies [47]. Our contribution is towards that aim, to make more accessible data experiences easier to design and implement within existing visualization work.

We present Data Navigator. Data Navigator is a toolkit built on a graph data structure, within which a broad array of common data structures can be expressed (including list, tree, graph, relational, spatial, diagrammatic, and geographic structures). Data Navigator also exposes an interface that supports interactions via screen reader, keyboard, gesture-based touch, motion gesture, voice, as well as fabricated and DIY input modalities. Data Navigator provides expressive structure and semantic rendering capabilities as well as the ability for developers to use their own, preferred method of rendering.

Data Navigator builds upon human-studies motivated work on accessible navigation [38, 47] towards a more generalizable resource for visualization practitioners. We contribute a high-level system design for our node-edge graph-based solution as well as an implementation of this system on the web, using JavaScript, HTML, and CSS. Through our case examples we also demonstrate that our generalized approach is suitable for replication of existing best practices from other systems, integration into existing visualization toolkit ecosystems, and development of novel prototypes for accessible navigation. We illustrate how Data Navigator's use of generic edges, dynamic navigation rules, and loose coupling between navigation and visual encodings provides practitioners robust, expressive, control over their system designs.

## 2 RELATED WORK

Our contribution is an attempt to bridge the gap between research and practice more effectively across broad ecosystems in order to enable deeper and more expressive accessible data navigation interfaces. Below we outline the prior research and standards that inform our project, a breakdown of existing visualization toolkit approaches to data navigation, and then accessible input device considerations.

### 2.1 Accessibility research and standards in visualization

Research and standards are both somewhat limited by a strong bias towards visual disabilities. In *Chartability*, 36 of the 50 criteria related to accessible visualization considerations involve visual disabilities [10, 11]. Marriott et al. also found that visual disability considerations are the primary focus of data visualization literature [27], leaving the barriers that many other demographics face unstudied.

However, despite the heavy focus on visual disabilities, the work that does exist in the visualization community is deeply valuable and serves as an important starting point for our technical contribution.

#### 2.1.1 Accessible navigation design considerations

Zong et al.'s research, which was conducted as in-depth co-design work and validated in usability studies involving blind participants, presented a design space for accessible, rich screen reader navigation of data visualizations. They organized their design space into *structure, navigation,* and *description* considerations and demonstrated example *structural*, *spatial*, and *direct* tree-based approaches [47].

*Chart Reader* also engaged these design space considerations in their co-design work on accessible data navigation structures [38]. We consider these design dimensions as the best starting point for our work, bridging the gap between research and toolkits.

There are additional research projects that have focused on accessible data navigation and interaction [14, 33, 34, 37]. These contributions explore a range of different interaction structures, including lists, trees, and tables of information as well as direct access methods such as voice interface commands and simple, pre-determined questions.

#### 2.1.2 Accessible visualization: understanding users

A wide array of emerging research projects investigate screen reader users needs, barriers, and preferences, and offer guidelines, models, and considerations for creating accessible data visualizations [4, 11, 25, 32]. Jung et al. offer guidance to consider the order of information in textual descriptions and during navigation [19]. Kim et al. collected screen reader users' questions when interacting with data visualizations, which could open the door for more natural language data interaction [20].

#### 2.1.3 Accessibility standards and guidelines

In the space of research, there has been a growing interest in developing guidelines for practitioners [8, 10] and even applying guidelines as a method of validation alongside human studies evaluations and co-design [11, 24, 25, 47]. Unfortunately, most accessibility standards and guidelines do not explicitly engage how to structure data navigation.

Despite this, existing accessibility standards bodies like the Web Content Accessibility Guidelines do stress the importance of accurate, functional semantics in order for screen reader users to know how to interact with elements [41]. For interactive visualizations this means that button-like or link-like behavior should expressly be made using elements that are semantically buttons and links. Our system should be capable of expressing meaningful semantics to users of assistive technologies.

### 2.2 Visualization toolkits and technical work

Unfortunately while many data visualization toolkits offer some degree of accessible navigation and interaction capabilities to developers, very few toolkits currently out there offer control over the important aspects of accessible data navigation design. Replicating existing research and strategies, remediating toolkit ecosystems, and building novel prototypes are all difficult or impossible to do due to the current lack of toolkit capabilities.

Existing data visualization toolkits have 3 major limitations that we wanted to address in the design of Data Navigator:

1. **Built on visual materials**: toolkits produce either raster or SVG-based visualizations, neither of which are focused towards designing navigable, semantic structures. As a consequence, many visualizations are simply entirely inaccessible.

2. **Lacking relational expressiveness**: When data navigation *is* provided, the navigation is based on either a tree or list structure (see Figure 2). The consequence of this limitation is that many other non-list and non-tree data relationships become difficult or impossible to represent without overly tedious navigation or inefficient architecture.

3. **Designed only for screen reader interaction**: When *accessible* data navigation is provided, it is generally only made possible through SVG with ARIA (Accessible Rich Internet Application) attributes. ARIA is primarily only leveraged by screen readers [42]. If a data element can be clicked and performs some form of function, only direct pointer (mouse and touch) and screen reader users are included. The consequence of this is that a wide array of other input devices, many used as assistive technologies by people with motor and dexterity disabilities, are excluded.

#### 2.2.1 Rich, tree-based approaches

De-coupling rendered, visual structures from meaningful and effective navigation experiences can provide richer experiences for screen reader users [47]. Prior research and industry work, with the exception of the *Visa Chart Components* library [39], has relied heavily on a 1 to 1 relationship between structure (the encoded marks) and navigation. This emerging work is significant, because it paves the way for considering the design dimensions of accessible data interaction and navigation without dependence on a visually encoded space.

*Olli's* approach has been to build ready-to-go adaptors that automatically build multiple tree structures for a few ecosystems (*Vega,*

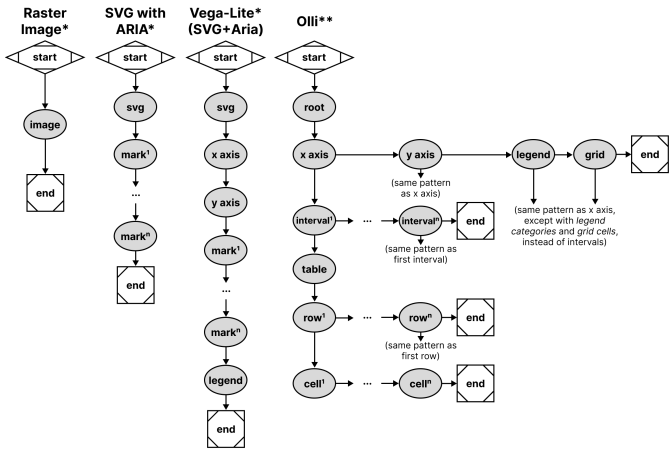

Fig. 2: Existing accessibility trees and lists, shown using node-edge graph conventions. (*) Denotes only *screen reader* access. (**) Denotes *screen reader*, *keyboard-only*, and *pointer* access as well.

*Vega-Lite*, and *Observable Plot*) and is entirely uncoupled from a data visualization's graphics. Their approach renders navigable tree structures *underneath* a visualization.

Other than *Olli*, *Highcharts* [16], *Visa Chart Components*, and *Progressive Accessibility Solutions'* visualization toolkits [14, 37] also primarily provide tree and list navigation structures across all of their chart types. These toolkits render their structures *upon* the visualization's graphic space. These tools also provide some degree of support for other assistive technologies and input modalities, although are limited exclusively to SVG rendering.

Unfortunately, these toolkits lack capabilities for dealing with graph, relational, spatial, diagrammatic, and geographic data structures.

#### 2.2.2 Serial, list-based approaches

Toolkits like *Vega-Lite* [31] and *Observable Plot* only provide basic screen reader support through ARIA attributes when visualizations are rendered using SVG. These libraries do not currently provide additional access to other assistive technologies and input modalities.

Microsoft's *PowerBI* largely uses a serial structure, although it has tree-like elements as well. *PowerBI* generally provides the same access to keyboard users as it does to screen readers, although not completely.

#### 2.2.3 No navigation provided

Other visualization tools, like *ggplot2* or *Datawrapper*, *Tableau*, as well as both *Vega-Lite* and *Highcharts* (when rendering to canvas), produce raster images and have no navigable structure available. Raster, or pixel-based graphics have been an accessibility burden since the early days of graphical user interface development [3]. Practitioners who use these toolkits can only provide alternative text.

### 2.3 Considering assistive technologies and input devices

Modern data visualizations may contain functional capabilities such as the ability to hover, click, select, drag, or perform some analytical tasks over the elements of the visualization space [31]. Virtually all of these analytical capabilities are designed for use with a mouse.

Input device consideration can roughly be organized as either *pointer-based* (such as a mouse or direct touch) or *non-pointer based* (which may employ speech recognition or sequential, discrete navigation such as with a keyboard). Assistive pointer-based devices, such as a head-mounted touch stylus, can typically perform any actions that a mouse can and are therefore served by current interactive visualizations. However, assistive non-pointer devices, such as a tongue, foot, or breath-operated switch, are not.

By only providing pointer-based interactivity, modern interactive visualizations exclude users who leverage non-pointer based input, who are most commonly people with motor and dexterity disabilities.

And unfortunately, there is a complete lack of engagement with these populations in the data visualization research community [27].

By comparison, the broader accessibility and HCI research communities have rich engagement with interaction and assistive technologies for users with motor and dexterity disabilities. Most research either focuses broadly on physical peripheral devices or sensors [36], wearables [30], or DIY making and fabrication [18].

The DIY making space involves a broad spectrum of complex input devices and materials, such as fabricating with wood and sensors for children with disabilities [22], 3D printed materials for rehabilitation professionals [13], and even using produce-based input (such as bananas and cucumbers) for aging populations [29].

Broadly, both research and practical developments related to accessible, non-pointer input are much further ahead than data visualization research and practice. Our goal for Data Navigator is to provide a technical resource towards engaging this under-addressed space.

## 3 DATA NAVIGATOR: SYSTEM DESIGN

We categorized our system design goals into design considerations for *Structure*, *Input*, and *Rendering*:

1. **Generic structure and navigation specification**: Human studies work has validated that lists, tables, trees, and even pseudo-treelike and direct structure types are all valuable to users in different contexts and with different considerations. Our system must be able to work with all of these as well as less frequently-used structures (spatial, relational, geographic, graph, and diagrammatic).

2. **Robust input handling**: Blind and low vision users may use combinations of different assistive technologies, such as magnifiers, voice interfaces, and screen readers. Users with motor impairments may rely on voice, gesture, eye-tracking, keyboard-interface peripherals (like sip-and-puffs or switches), or fabricated devices. Both the developer and user should therefore be able to leverage and customize a broad range of input types, including the above as well as fabricated, adaptive, and future input modalities.

3. **Flexible rendering and semantics**: Visuals may or may not be necessary to render to demonstrate Data Navigator's structure. In addition, much of the latest research has shown that different screen reader users may prefer different orders of information and at different levels of verbosity. In addition, the context of tasks the user is performing as well as the nature of the data itself may influence the design of semantic descriptions and visual indications for elements. Data Navigator must provide a high degree of flexibility and control.

To help bridge the gaps between research and standards knowledge about best practices and building an effective toolkit for practitioners, we intend for Data Navigator to provide both *exploratory support* and *vocabulary correspondence* [26].

In particular, our ideal users are developers who specify data visualizations using code. To that aim, we intend to provide *exploratory support* through generic, dynamic, and flexible system design decisions. Our system is expressive and customizable, which encourages exploration of different options.

And we also want the API to include properties that have conceptual and *vocabulary correspondence* to our design considerations. Each design consideration (*Structure*, *Input*, and *Rendering*) are separately composable, modular subsystems of Data Navigator that can be used independently or in tandem with one another.

In this paper we present an implementation of our system using JavaScript, HTML, and CSS on the web. The demonstration of our system is best suited to the web due to the nature of existing, accessible building blocks (HTML), which resolve many of the semantic complexities and logic involved in enabling screen readers to programmatically navigate and announce meaningful information to users. In addition, many existing visualization toolkits target the web as an output platform and we believe that this is the best starting point for adoption and use of Data Navigator. However, this system design could be implemented as a toolkit in other environments with proper consideration for input device handling and screen reader semantics.

## 3.1 Structure

### 3.1.1 Beyond trees: towards an accessibility *graph*

The first major contribution in the design of Data Navigator is to use node-edge data as the substrate for our navigation system.

The most important argument in favor of using a graph-based approach is that a graph can construct virtually any other data structure type (see Figure 2), including list, table, tree, spatial, geographic, and diagrammatic structures. Graphs are generic, which enables them to represent structures both in current and future interface practices [12].

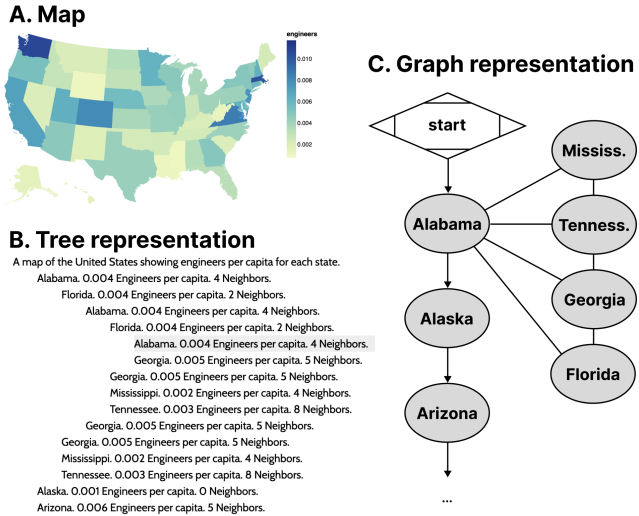

Fig. 3: **A.** Map of engineers per capita of US states. **B.** Tree representation of the map data where states are listed alphabetically and also include links to neighboring states. The structure repeats itself if users navigate in a loop. **C.** Graph representation with the same navigation potential without redundant rendering.

To demonstrate our point, the most recent emerging work with advancements in accessible data navigation used node-edge diagrams to demonstrate their tree-like structures [38, 47] similar to Figure 2, Figure 9, and Figure 10. This is because trees are a form of node-edge graph, but with a root, siblings, parents, and children as sub-types of nodes that generally have rules for how they relate to one another.

Node-edge graph structures prioritize direct relationships. Examples of common direct relationships in visualization are boundaries on maps (see Figure 3), flows and cycles, data with multiple high level tree structures pointing to the same child datasets (such as *Olli* in Figure 2), or even just in diagrammatic, graph-based visualizations.

A graph structure allows for direct access between information elements that are not just part of the input data or 1:1 rendered elements, but may also have perceptual or human-attributed meaning. Examples of this might include semantic or task-based relationships, such as navigating to annotations or callouts, between visual-analytic features like trends, comparisons, or outliers. Spatial layouts such as intersections of sets or parallel vectors (see Section 4.3), or even relationships to information outside of a visualization and back into it (like in Figure 7) are enabled by a graph structure.

### 3.1.2 Graph structures are more computationally efficient

Data visualizations often portray information that becomes difficult to handle when using trees and lists. The distance users must travel between relational elements is significant in lists while redundancy when navigating relational elements in trees can be problematic.

As an example of this, often a data table or list of locations are used in conjunction to a map, such as listing all 50 states alphabetically along with relevant information. The list itself is expensive to navigate and may not provide any relationship information about which states border others, let alone ways to easily and directly access those states.

Part of the visual design justification of using a map instead of a table is for sighted individuals to understand how geospatial information may interact with a given variable. The spatial relationships matter. But when supplementing the list of states with sub-lists for each state's bordering states (see Figure 3), it produces redundancy in the rendered result. The rendered data contains circular connections between nodes but must render every reference, producing a computational resource creep and cluttered user experience that can be difficult to exit.

### 3.1.3 Specific edge instances and generic edges

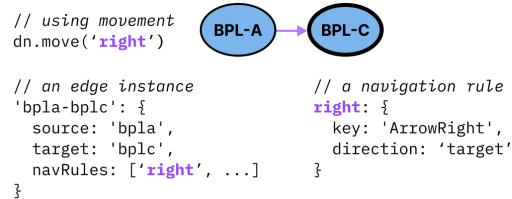

```
// using movement
dn.move('right')

// an edge instance              // a navigation rule
'bpla-bplc': {                   right: {
  source: 'bpla',                  key: 'ArrowRight',
  target: 'bplc',                  direction: 'target'
  navRules: ['right', ...]       }
}
```

Fig. 4: An example of how a single edge instance references a navigation rule and can even have multiple navigation rules. A navigation rule can be referenced by multiple edges.

In Data Navigator, nodes are *objects* that always contain a set of edges, where each edge contains a minimum of 4 pieces of information: a unique identifier, a source, a target, and navigation rules. These properties are only accessed when a navigation event occurs on a node with an edge that contains a reference to a rule for that navigation event. Navigation rules may be unique to an edge instance or shared among other edge instances.

The source and target properties of edges are either ids that reference node instances (see Figure 4) or *functions* (see Figure 5). Because some edges in a graph may be directed or not, non-directed graphs can use source and target properties to arbitrarily refer to either node attached to an edge.

Generic functions for source or target properties can link nodes to other nodes based on changing content, structure, or behavior that may be difficult or impossible to determine before a user navigates the structure.

Function calling also allows some edges to be *purely* generic. An example of a reasonable use case of a purely generic edge is in Figure 5, where the source is a function which returns the present node and the target is whichever node the user was on previously. This single edge may then be part of every node's set of edges, enabling users to have a simple *undo* navigation control without creating an *undo* edge unique to every source node.

Using this pattern, it is possible to have fully navigable structures using only generic edges.

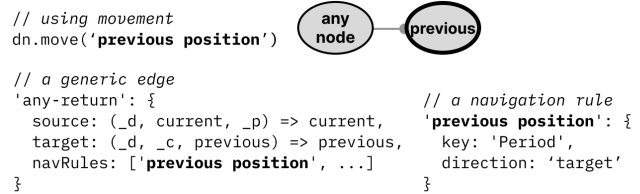

```
// using movement
dn.move('previous position')

// a generic edge                        // a navigation rule
'any-return': {                          'previous position': {
  source: (_d, current, _p) => current,    key: 'Period',
  target: (_d, _c, previous) => previous,  direction: 'target'
  navRules: ['previous position', ...]   }
}
```

Fig. 5: A generic edge, such as "any-return" can be applied to any node. Function calls handle dynamically assigning the edge's source and target nodes on-demand.

## 3.2 Input

### 3.2.1 Abstracted navigation facilitates agnostic input

Navigation rules in Data Navigator (see Figure 4 and Figure 5) are created alongside the node-edge structure. Edges reference rules for navigation. However, these rules are generic and agnostic to the specifics of input modalities and can be invoked as methods by virtually any detected user input event (see Figure 6).

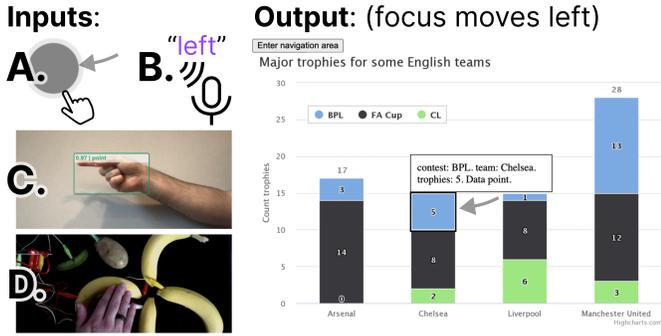

Fig. 6: An example navigation rule to move "left" can be called as a method by an event from any input modality. Some examples include common modalities such as touch swiping (**A**) or speaking "left" (**B**). This also includes advanced or future modalities such as gesture recognition (**C**) or touch-activated, fabricated interfaces (**D**).

Navigation rules are objects with a unique name, ideally as a noun or verb in natural language that refers to a direction or location, a movement direction (a binary used to determine moving towards the source or target of an edge), and optionally any known user inputs that activate that navigation, such as a keyboard keypress event name.

It is important for a system to abstract navigation events so that inputs can be uncoupled from the logic of Data Navigator. This allows higher level software or hardware logic to handle input validation while Data Navigator is just responsible for acting on validated input.

Later in our first case example (Section 4.1), we demonstrate an application that handles screen reader, keyboard, mouse and touch (pointer) swiping, hand gestures, typed text, and speech recognition input. Abstract navigation namespaces can be called by any of these input methods.

Additionally, since navigation rules are flexible, end users can also supply their own key-bind remapping preferences or input validation rules if developers provide them with an interface.

Because calling a navigation method is abstract, users can even supply events from their own input modalities as long they have access to either a text input interface or access to Data Navigator's navigation methods. Our demonstration material (in Section 4.1) also includes handling for DIY fabricated interfaces, which are important in accessibility maker spaces. We chose a produce-based interface [29], since it was an easy and low cost proof of concept.

We believe that enabling agnostic input provides a rich space for future research projects. In addition, browser addons and assistive technologies could both leverage this flexible interface for end users.

#### 3.2.2 Discrete, sequential input opens new avenues

The *keyboard interface* is considered foundational for many assistive devices, which leverage this technology for discrete, sequential, non-pointer navigation and interaction [43]. Desktop screen readers are the most common example of an assistive technology device that leverages the keyboard interface, however single or limited button switches, sip-and-puff devices, on-screen keyboards, and many refreshable braille displays do as well. Support for the keyboard interface by default in turn provides all discrete, sequential input devices with access as well.

However by basing Data Navigator's foundational infrastructure on a keyboard-like modality, this also provides designers and developers new avenues to imagine how existing direct, pointer-based, or continuous inputs can map to discrete, sequential navigation experiences.

For example, with mobile screen readers this already happens: screen reader users swipe and tap on their screen to sequentially navigate, but the exact pixel locations of their swiping and tapping generally does not matter. Their current focus position is discrete and determined by the screen reader software.

Data Navigator therefore allows for many new possibilities. One possibility is that sighted mouse and touch users may now also swipe their way through dense plots or use small interfaces (such as on mobile devices) that may otherwise be too hard to precisely tap. Data Navigator optionally removes the accessibility barriers sometimes posed by precision-based input in visualizations.

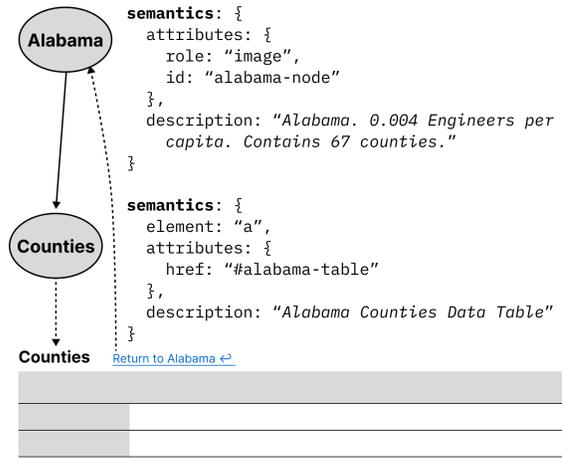

Fig. 7: An example of how navigation within Data Navigator could use semantic nodes as hyperlinks to provide access to other areas in an application. Alabama has a child node "Counties" which is a semantic HTML link element pointing to a table of counties, outside of Data Navigator's graph structure. A link is provided to return.

Data Navigator does not have to be in conflict with precision-based input, either. A discrete, sequential navigation infrastructure can be used in tandem with precision-based pointer events as well as instant access when coupled with voice commands and search features.

### 3.3 Rendering

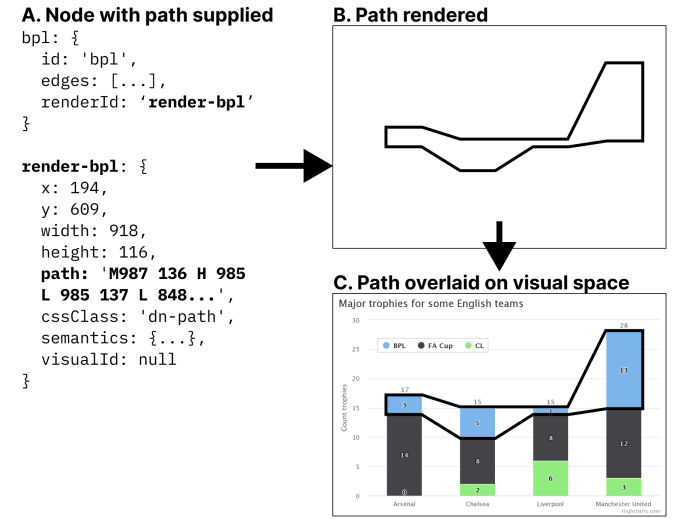

Fig. 8: **A.** The data specified for a node with a reference to separate data that is used to render that node. **B.** The node will render as a path at the specified Cartesian coordinates. **C.** This rendered node may then be placed over a visual.

#### 3.3.1 Flexible node semantics provide freedom

Nodes in Data Navigator are semantically flexible. This is because the marks in a data visualization may represent many things, that are either dependent on the data or the user interface materials.

Since our toolkit implementation is in JavaScript and HTML, our map example from Figure 3 might use image semantics for states,

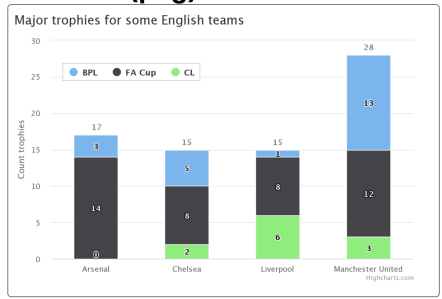
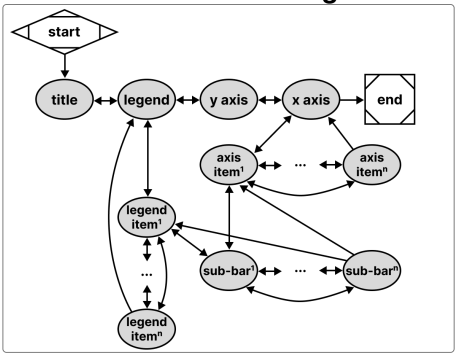
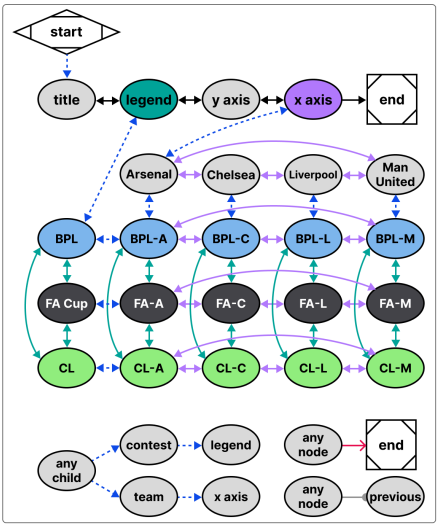
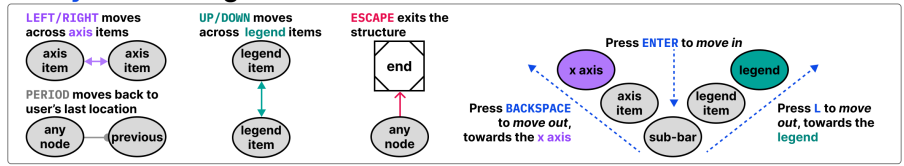

Fig. 9: **A.** A raster (png) visualization of a stacked bar chart showing how 4 English teams performed across 3 major trophy contests. **B.** An example navigation schema that allows children nodes to have 2 parents (two tree structures intersecting), one for contests and one for teams. **C.** An example of Data Navigator's navigation logic abstraction, which allows edge types to have programmatic sources, targets, and rules, such as a single rule that gives all nodes a edge to exit the visualization. **D.** An instantiation of the schema, showing all corresponding rendered nodes and their edge types according to the schema design and navigation rules.

alongside a description of the data relevant to that node. However since semantics are flexible in this way, Data Navigator could also be used to integrate into a larger ecosystem, with nodes rendered as hyperlinks to tables or other elements such as in Figure 7.

The concept of using node-edge graphs can even extend to have "nodes" that are entirely different parts of a document or tool, as well as integrated into the explicit structure provided by Data Navigator. In some accessibility toolkits, nodes are geometries without functional semantics [31] or list items nested within lists [1]. But in Data Navigator, nodes can semantically be buttons, links, or any HTML element. Interactive data visualizations sometimes demand more flexible node semantics than geometries or lists.

### 3.3.2 Loose-coupling to visuals enables expressiveness

One of the most significant technical limitations of existing data visualization toolkits with regards to accessibility is that they rely on visual substrate, or visual materials, in order to produce data visualizations. In the case of static, raster images such as png files or WebGL and canvas elements on the web, there are no interface properties at all exposed to screen readers for programmatic exploration and interaction.

If raster images are used, they generally cannot be changed after rendering. However, according to web accessibility standards, elements must have a visual indicator provided when focused [40].

Since Data Navigator navigates using focus, an indicator must be rendered alongside the node semantics. But *what* is focused visually and *where* it is depends on different design needs.

In *Visa Chart Components*, chart elements can be *selected*, so the focus indication is visible over the existing elements in the chart space. The design choice to have interactive visual elements located within a chart or graph is also common in other toolkits that provide accessible focus indication, such as *Highcharts*, *PowerBI*, and *SAS Graphics Accelerator*.

However, some visualization toolkits create accessible structures entirely uncoupled from visual space [1], so focus indication is provided beneath or beside the chart, not over it.

Due to the different ways that accessibility might be provided, Data Navigator enables developers to have complete control over the rendering of which focus elements they want, in what styling, and where. This can accommodate both un-coupled and visually-coupled approaches to focusing and more.

Data Navigator's focus is *uncoupled* by default and may even be used independent of any existing graphics at all. Rendering information may be passed to Data Navigator for it to render (like in Figure 8) or developers can provide their own rendered elements and simply use Data Navigator to move between them.

Because of Data Navigator's approach to rendering focusable elements, designers and developers can provide fully customized annotations, graphics, text, or marks that may not be not part of the original visual space or elements. One example of this might be adding an outlined path to a collective cross-stack group of bars in a stacked bar chart (see Figure 8).

Loose-coupling in this way provides robust flexibility to designers and developers to handle navigation paths and stories through a data visualization, even in bespoke or hand-crafted ways.

### 3.3.3 On-demand node rendering is efficient

Practitioners care about performance and so do users. Practitioner toolkits often focus on lazy-loading techniques where accessibility elements are rendered on-demand rather than all in-memory up front [1, 9, 47].

Data Navigator's nodes are rendered *on-demand* by default. Data Navigator only renders the node that is about to be focused by the user and after it is focused, the previously focused node is deleted from memory. This technique has advantages in cases where datasets are large or users have lower computational bandwidth available. However, there are cases where practitioners may want to render all of Data Navigator's structure in memory, such as server-side rendering or equivalent. Pre-rendering may be optionally enabled.

## 4 CASE EXAMPLES WITH DATA NAVIGATOR

We built example prototypes using our JavaScript implementation of Data Navigator, available open source at our GitHub repository.

Our first two prototype case examples represent some of the most powerful parts of Data Navigator as a system while reproducing known and effective data navigation patterns from existing industry and research projects. We provide a final case example as a co-design session that demonstrates how Data Navigator may be used to rapidly build new designs.

### 4.1 Augmenting a Static, Raster Visualization

The first case example (shown in Figure 9) builds on an online JavaScript visualization library, *Highcharts*. *Highcharts* already provides relatively robust data navigation handling out of the box for screen reader, keyboard, and even voice recognition interface technologies, such as *Dragon Naturally Speaking*. However, these capabilities are only provided when the chart is rendered using SVG. Developers have several other rendering options available, including WebGL, which is significantly more efficient [15]. We wanted to demonstrate that Data Navigator can provide a navigable data structure even if the underlying visualization is a raster image.

For our case example, we exported a png file using the built in menu of a sample stacked bar chart retrieved from their online demos [17]. We selected a stacked bar chart because it allows us to demonstrate how two tree structures may interact and share the same children nodes.

We recorded the data and hand-created all of the geometries and their spatial coordinates using *Figma*, by tracing lines over the raster image's geometries (see samples of the data and traced geometries in Figure 8). While this method was efficient for building an initial prototype, Section 4.2 engages deterministic methods for extracting and producing the nodes, edges, and descriptions required by Data Navigator automatically and at scale.

The visualization we selected represents 4 English football teams, *Arsenal, Chelsea, Liverpool,* and *Manchester United* and how many trophies they won across 3 contests, *BPL, FA Cup,* and *CL*.

We chose a schema design that arranged the *contests* to be navigable across one dimension of movement (*up* and *down*) while the *teams* are navigable across a perpendicular dimension of movement (*left* and *right*). This 2-axis style of navigation is used by *Highcharts* (when rendering as SVG) and *Visa Chart Components*. We also chose these directions because it is coincidental that their visual affordance is closely coupled with the navigation design (the x axis is ordered *left* to *right* and since the bars are stacked, *up* and *down* can move within the stack). These directions can also be applied to the axis categories and legend categories as well, moving *left* and *right* across the entire *team's* stacks or up and down across the entire *contest's* groupings.

Using a keyboard, a user might enter this schema and navigate to the legend, where they could press *Enter* to then focus the legend's first child, pictured in Figure 8. Pressing up or down navigates in a circular fashion among the *contest* groupings. Pressing *Enter* again then focuses the first child element of that *contest*, all of which are in the *Arsenal* group, since it is the first group along the x axis. A user can then navigate *up, down, left,* and *right* among children. Pressing *L Key* moves the user back up towards the contest while pressing *Backspace* moves the user up towards the x axis. The x axis and *team* groupings represent the second tree which intersects the first (the *contests*).

Our first case example includes handling for additional input modalities beyond screen readers and keyboards, including a hand gesture recognition model, swipe-based touch navigation, and text input (which can be controlled using voice recognition software).

#### 4.1.1 Discussion

Our first case example demonstrates several of the most important capabilities of Data Navigator, namely that practitioners can add accessible navigation to previously inaccessible, static, raster image formats and that a wide variety of input modalities are supported easily.

Widely-used toolkits like *Vega-Lite*, *Highcharts*, and *D3* [2] allow practitioners to choose SVG and canvas-based rendering methods. Data Navigator's affordances help overcome the lack of semantic structure in canvas-based rendering, allowing developers to take advantage of its processing and memory efficiency.

Notably in addition to these capabilities, the visual focus highlighting added was entirely bespoke (as in Figure 8) and the navigation paths through the visual were based on our design intentions, not an extracted view or underlying architecture such as render order. This demonstrates that our system provides a significant degree of freedom and control for designers and developers.

As a final discussion point, the resulting visualization contains no automatically detectable accessibility conformance failures according to the W3C's Web Accessibility Initiative's accessibility evaluation tool, *WAVE* [44]. It is important for any technology developed to also meet minimum requirements for accessibility [10, 24, 25, 47], even when following best-practices and research.

### 4.2 Building Data Navigation for a Toolkit Ecosystem

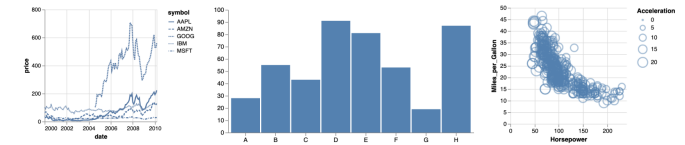

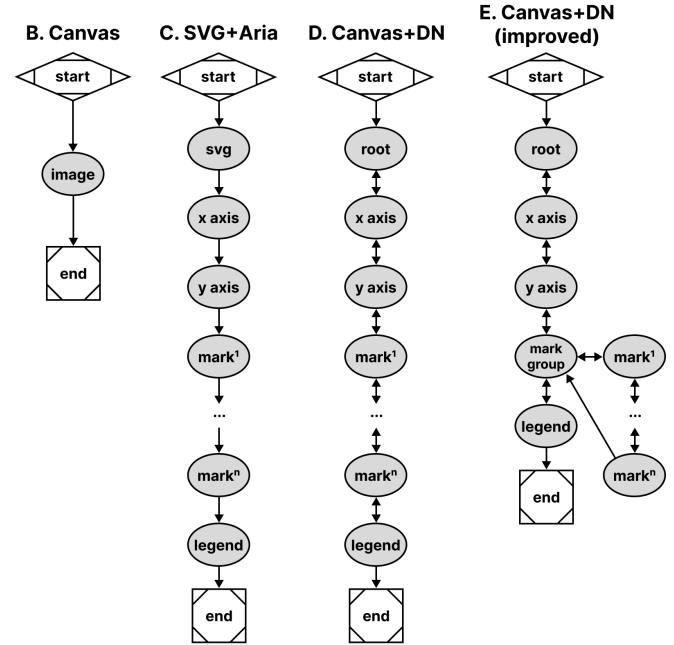

Fig. 10: **A.** Various charts from *Vega-Lite* share the same general structures with each other when rendered using canvas (**B**) or SVG (**C**). **D.** With Data Navigator, we replicated the existing SVG navigation pattern (**C**) but used a canvas-based rendering for the visualization. **E.** We also improved the navigation scheme to nest marks within a mark group to allow users to skip them, if needed.

Our second case example, shown in Figure 10, builds on *Vega-Lite*. As shown in Figure 2, *Vega-Lite* offers basic screen reader navigation but provides no navigation at all when rendered using canvas.

While it might be a tedious design choice to allow every mark in a visualization to be serially accessible to screen reader users, we nevertheless set out to build a generic ingestion function that would take a *Vega-Lite View* object and deterministically recreate their existing SVG navigation structure in Data Navigator. This way users would have the same experience between SVG rendered charts and all current and future rendering options that *Vega-Lite* offers to developers.

Notably, *Vega-Lite* does not explicitly manipulate the navigation order at all when rendering with SVG. ARIA is simply provided to allow screen reader users to access each mark in the visualization in the order the mark appears in the DOM (which is the order it was rendered). The legend appears after the marks in our schema for this reason because *Vega-Lite* renders the legend after marks. This choice of ordering is for visual reasons: z-axis placement is currently based on render order in SVG and *Vega-Lite* wants their legend visually on top of the rendered marks.

In addition to mimicking their existing SVG navigation strategy, we also created a way to nest all of the marks within a group so that users can skip past them and drill in on-demand, which is a valuable pattern when dealing with situations where providing a mark-level fidelity of information may not be relevant to a user's needs by default [35, 47].

In order to deterministically supply Data Navigator with accurate information about any given *Vega-Lite* visualization, we built 3 functions: one that takes a *Vega-Lite View* as input and extracts meaningful nodes, one that produces edges based on those nodes, and one to describe our nodes in a meaningful way for screen reader users. These generic functions technically work on all existing *Vega-Lite* charts, however some are more useful out of the box than others due to the type of marks involved.

### 4.2.1 Discussion

This case example demonstrates that ecosystem-level remediation and customization is not only possible for toolkit builders but Data Navigator offers robust potential. Data Navigator's structure, input, and rendering capabilities are all flexible and can be adjusted to suit the needs of a specific toolkit's design and intended use.

Many visualization libraries may not even provide screen reader accessible SVG using ARIA-based approaches but do have a consistent underlying architectural pattern. Some libraries have a consistent method for converting data into visual formats, readable text labels, and interaction logic. Strong contenders would be visualization libraries popular in online, web-based data science notebooks like *ggplot2* in R or *matplotlib* for Python, which typically only render rasterized pngs or semanticless SVG.

Toolkits with consistent underlying architecture would allow toolkit developers, not just developers who *use* toolkits, to remediate and customize their navigation accessibility using a generic approach.

Enabling accessibility at the toolkit level allows all downstream use of that tool to have better defaults, options, and resources available for building more accessible outcomes for end users.

Many libraries and toolkits provide users with a level of functional defaults and abstract conciseness so that users don't have to worry about low-level geometric considerations [31].

Data Navigator allows toolkits developers to also provide their users with abstractions and defaults for accessibility that make sense for their ecosystem.

Despite our schema recreating a screen reader experience based on SVG (and improving it), Data Navigator's additional features also apply: users are able to leverage a much wider array of input modalities.

*Vega-Lite* provides many ways to make marks clickable and even perform complex actions using mouse-based input. While Data Navigator does not engage accessible brush and drag-based inputs, it does provide keyboard-only access by default, which can be used to make events previously only accessible to mouse clicking available to many other technologies. This is an improvement over *Vega-Lite*'s SVG + ARIA rendering option.

When measuring performance across test datasets containing 406 and 20,300 data points in a scatter plot, Data Navigator increases initialization time by ∼0.45 to ∼1.5ms respectively. Our extraction functions specific to *Vega-Lite* increase initialization between ∼4.8 and ∼8.5ms respectively. Given that our benchmark testing for *Vega-Lite*'s SVG rendering initialized in ∼1,800ms for 20,300 data points and canvas in ∼700ms, we do not anticipate that Data Navigator will have a negative impact on performance in most visualization contexts.

## 4.3 Co-designing Novel Data Navigation Prototypes

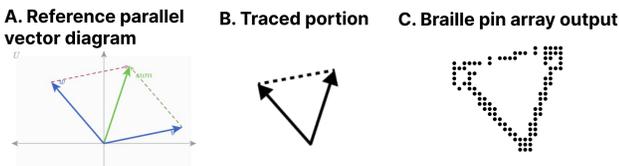

Fig. 11: Our material preparation process involved taking a reference (**A**), tracing it (**B**), and rendering it on a tactile display (**C**).

Recent projects in accessible data navigation have involved extensive co-design work with people with disabilities, ranging on the magnitude of months with as many as 10 co-designers at a time [24, 25, 38, 47].

However many visualization experiences may be authored in smaller scales, with fewer designers, and less time such as the development of a prototype or demonstration of an emerging idea. In practical or industry contexts, co-design sessions (and design sessions in general) may be much shorter. The goal of these co-design sessions is simply to create an artifact with the artifact's intended users.

Since our paper is contribution towards practical outcomes, we simulated a light co-design session with the aim of producing low-fidelity prototypes of novel data interaction patterns.

### 4.3.1 Co-design Session Methods and Setup

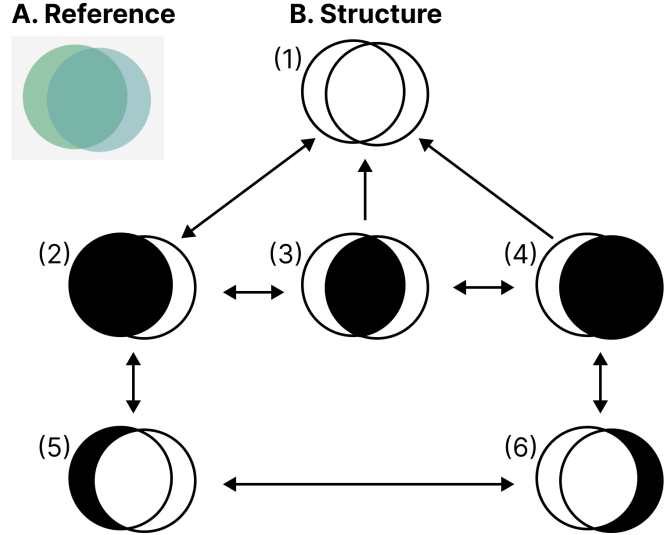

Fig. 12: **A.** A reference image from *Penrose* of a set diagram containing two sets intersecting. **B.** A diagram of our proposed structure, with three levels of information.

Authors Frank Elavsky (sighted) and Lucas Nadolskis (blind) set out with the goal of developing screen-reader friendly prototypes that can explore geometric and mathematical models produced by the math diagramming tool *Penrose* [46].

Nadolskis is a neuroscience engineer who is a native screen reader user and uses both mathematical concepts as well as data-related tasks in his research. Elavsky proposed a series of possible math-based visualization types produced by *Penrose* to build prototypes for, and Nadolskis selected *set* and *vector* diagrams as the two worth exploring first. The justification for this selection is that understanding these two concepts is important for work in data science, programming, and more advanced math concepts.

In particular we grounded the context of our contribution in a hypothetical classroom setting, where a screen reader user who is a student will have access to the equations in both raw text and *MathJax*. We want to provide an experience that does not replace the existing resources screen reader users have to learn in classrooms but rather supplement.

At our disposal for our co-design session was a *Dot Pad* [6], which is a refreshable tactile braille display. Our *Dot Pad* enabled Elavsky to produce something visual and then translate it into the display for Nadolskis. Similar to de Greef et al. [5], we used a tactile interface as an intermediary to help us get a shared sense of the meaningful spatial features of our figures.

Elavsky started with a reference diagram and then traced a wide variety of every possible node that might be worth navigating to in the diagram (see Figure 11).

We selected which nodes were most important in each diagram, how to navigate between them, and how we wanted to render their visuals and semantics.

The selection of our problem space, scope of solutions, context of contribution, general discussion, and preparation of materials took approximately 12 hours of work over 2 weeks. The exploration of our prototype design space for our 2 prototypes took 1 hour. Building the prototypes took 2 hours.

### 4.3.2 Creating a Navigable Set Diagram

Our first prototype was a set diagram (see Figure 12). For our structure, we decided that it has 3 important semantic levels: the high level, the inclusion level, and the exclusion level. The inclusion level is first and the siblings are all sets or subsets that include other sets. The exclusion level is beneath and contains sets or subsets that are exclusive to the sets they belong to, which are accessed by drilling down from a set.

Our schema design starts with a user encountering the root level (1) and may optionally drill in to the first child of the next level (2) using the *Enter* key. The user may navigate siblings at this level using *right* and *left* directions, but this level is not circular (like in Figure 9) to maintain the spatial relationships. The user may drill in on either set again to view the non-intersecting portion of that set. Any node can drill up, towards the root, using *Escape* or *Backspace*.

### 4.3.3 Creating a Navigable Parallel Vectors Diagram

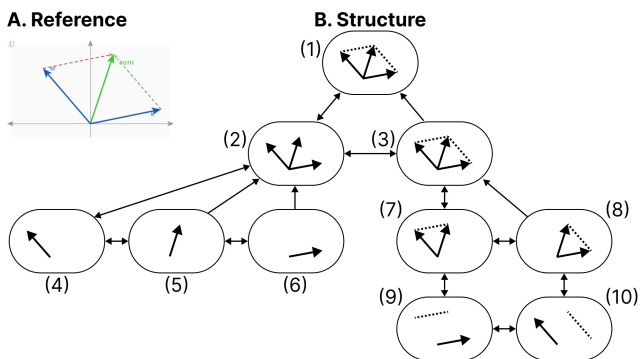

Fig. 13: **A.** A reference image from *Penrose* of a parallel vectors diagram. **B.** A diagram of our proposed structure, with two main sub-categories of information: understanding the vectors and their parallels.

Our second prototype was a parallel vectors diagram (see Figure 13). For the structure of this diagram we created a first level group that contains each vector and vector sum. The sibling to this grouping is another group which organizes sub-equations related to calculating each parallel vector. The sub equations each contain children that pair the sub equation with the vector it is parallel to.

Similar to Figure 12, this figure maintains spatial relationships along the x dimension, does not have circular navigation, and allows drilling in and out.

### 4.3.4 Discussion

After our co-design sessions, our visual materials and navigation structures were used in the creation of functional prototypes. We additionally hand-crafted the descriptions and semantics for each node.

Accessibility work often takes a long time, from co-design to building to validation. But we believe that a well-articulated and useful design space, with tools that provide expressiveness and control over the dimensions of that design space, can improve how this work is done. The above case example demonstrates how builders who are thinking about data navigation design can rapidly scaffold prototypes for use in Data Navigator.

In particular, Data Navigator's design as a system gave our co-design sessions *vocabulary correspondence*. Data Navigator's language helped us focus on the *nodes, edges,* and *navigation rules* for our *structure* while we also explicitly discussed the *rendering* details of *coordinates, shapes, styling,* and *semantics* for each node. The vocabulary of our design space directly corresponded with code details required to create a functional prototype.

We note that this co-design work is not intended to contribute a *validated* set of designs. Rather, our contribution with this case example is to demonstrate that within the larger ecosystem of a research venture, Data Navigator is an improvement over designing and building navigable structures from scratch.

## 5 LIMITATIONS AND FUTURE WORK

Data Navigator is a technical contribution, a system designed for appropriation [7] and adaptation [45] in different applied contexts. It is, as Louridas writes, a *technical material*: a technology that enables new and useful capabilities [23]. While beyond the scope of the current paper, a critical next step for future work is to conduct separate studies with both practitioners and end users to evaluate Data Navigator's affordances.

Unlike toolkits that provide an end-to-end development pipeline for accessible visualization, Data Navigator serves as a low-level building block or material (like concrete). As such, one potential limitation of the framework is that it can be used to build both curbs (which are inaccessible) as well as ramps and *curb-cuts* (which may be more broadly accessible).

Even when building more accessible curb-cuts, we stress the importance of actively involving people with disabilities in the design and validation of new ideas, in line with prior work [24, 25, 28, 47]. For example, while our first two case examples replicate co-designed and validated existing work, our third case example's co-designed prototypes would need to be validated with relevant stakeholders before wider implementation. Our system does not *guarantee* any sort of accessibility on its own.

The diverse array of modalities supported by Data Navigator opens an immediate line of future work in engaging people with a correspondingly diverse set of disabilities. While recent explorations into accessible data visualization have been inspiring, this trend has primarily focused on the experiences of people with visual disabilities [10, 21, 27]. More research should be conducted with other populations, particularly people who leverage assistive technologies beyond screen readers, to understand how interactive data visualizations can be better designed to serve them.

Finally, there are significant opportunities to improve the efficiency of our approach, including developing deterministic and non-deterministic methods to generate node-edge data and navigation rules from a visualization. Ma'ayan et al. stress in particular that reducing tedious complexity can contribute to the success of a well-designed toolkit [26]. Future work should identify areas where graphical interface tools or higher-level specifications can improve the experience of working with Data Navigator.

## 6 CONCLUSION

Practitioners at large continue to produce inaccessible interactive data visualizations, excluding people with disabilities. We believe that the burden of remediation first starts with the developers who build and maintain the toolkits that practitioners use.

However, the challenges faced by toolkit builders are significant. Most toolkits lack an underlying, navigable structure, support for broad input modalities used by people with disabilities, and meaningful, semantic rendering.

To engage these limitations we present Data Navigator, a technical contribution that builds on existing work towards a more generalizable accessibility-centered toolkit for creating data navigation interfaces. Data Navigator is designed for use by practitioners who both build and use existing toolkits and want a tool to make their data visualizations and interfaces more accessible.

We contribute a high-level system design for our node-edge graph-based approach that can be used to build data structures that are navigable by a wide array of assistive technologies and input modalities. Data Navigator is generic and can scaffold list, tree, graph, relational, spatial, diagrammatic, and geographic types of data structures common to data visualization.

Our system is designed to encourage both remediation of existing inaccessible systems and visualization formats as well as help scaffold the design of novel, future projects. We look forward to further research that explores the possibilities enabled by Data Navigator.

## ACKNOWLEDGMENTS

We want to take this time to express immense gratitude for Reviewer 1, whose generous and thorough feedback helped this project find its true vision. Elavsky also wants to thank the many folks who have encouraged this project's ideation and formation over the last few years.

This work was supported by a grant from Apple, Inc. Any views, opinions, findings, and conclusions or recommendations expressed in this material are those of the authors and should not be interpreted as reflecting the views, policies or position, either expressed or implied, of Apple Inc.

## REFERENCES

[1] M. Blanco, J. Zong, and A. Satyanarayan. Olli: An extensible visualization library for screen reader accessibility. In *IEEE VIS Posters*, 2022. 6

[2] M. Bostock, V. Ogievetsky, and J. Heer. D3 data-driven documents. *IEEE Transactions on Visualization and Computer Graphics*, 17(12):2301–2309, Dec. 2011. doi: 10.1109/TVCG.2011.185 7

[3] L. Boyd, W. Boyd, and G. Vanderheiden. The graphical user interface: Crisis, danger, and opportunity. *Journal of Visual Impairment & Blindness*, 84(10):496–502, Dec. 1990. doi: 10.1177/0145482x9008401002 3

[4] P. Chundury, B. Patnaik, Y. Reyazuddin, C. Tang, J. Lazar, and N. Elmqvist. Towards understanding sensory substitution for accessible visualization: An interview study. *IEEE Transactions on Visualization and Computer Graphics*, 28(1):1084–1094, Jan. 2022. doi: 10.1109/tvcg.2021.3114829 2

[5] L. de Greef, D. Moritz, and C. Bennett. Interdependent variables: Remotely designing tactile graphics for an accessible workflow. In *The 23rd International ACM SIGACCESS Conference on Computers and Accessibility*. ACM, Oct. 2021. doi: 10.1145/3441852.3476468 8

[6] Dot Pad inc. Dot pad - the first tactile graphics display for the visually impaired. https://pad.dotincorp.com/, 2020. 8

[7] P. Dourish. The appropriation of interactive technologies: Some lessons from placeless documents. *Computer Supported Cooperative Work (CSCW)*, 12(4):465–490, 2003. doi: 10.1023/a:1026149119426 9

[8] E. Durant, M. Rouard, E. W. Ganko, C. Muller, A. M. Cleary, A. D. Farmer, M. Conte, and F. Sabot. Ten simple rules for developing visualization tools in genomics. *PLOS Computational Biology*, 18(11), Nov. 2022. doi: 10.1371/journal.pcbi.1010622 2

[9] F. Elavsky. Method and system for accessible data visualization on a web platform. *Defensive Publication Series*, 4220, Apr. 2021. 6

[10] F. Elavsky, C. Bennett, and D. Moritz. How accessible is my visualization? evaluating visualization accessibility with chartability. *Computer Graphics Forum*, 41(3):57–70, June 2022. doi: 10.1111/cgf.14522 2, 7, 9

[11] D. Fan, A. F. Siu, H. Rao, G. S.-H. Kim, X. Vazquez, L. Greco, S. O'Modhrain, and S. Follmer. The accessibility of data visualizations on the web for screen reader users: Practices and experiences during COVID-19. *ACM Transactions on Accessible Computing*, 16(1):1–29, Mar. 2023. doi: 10.1145/3557899 2

[12] E. R. Gansner and S. C. North. An open graph visualization system and its applications to software engineering. *Software: Practice and Experience*, 30(11):1203–1233, 2000. doi: 10.1002/1097-024x(200009)30:11<1203::aid-spe338>3.0.co;2-n 4

[13] S. Giraud and C. Jouffrais. Empowering low-vision rehabilitation professionals with "do-it-yourself" methods. In *Lecture Notes in Computer Science*, pp. 61–68. Springer International Publishing, 2016. doi: 10.1007/978-3-319-41267-2_9 3

[14] A. J. R. Godfrey, P. Murrell, and V. Sorge. An accessible interaction model for data visualisation in statistics. In *Lecture Notes in Computer Science*, pp. 590–597. Springer International Publishing, 2018. doi: 10.1007/978-3-319-94277-3_92 2, 3

[15] Highsoft. Highcharts: Render millions of chart points with the boost module. https://www.highcharts.com/blog/tutorials/highcharts-high-performance-boost-module/. Accessed: 2022-11-20. 7

[16] Highsoft. Highcharts accessibility module: information and demos. https://highcharts.com/docs/accessibility/accessibility-module, 2018. Accessed: 2021-09-06. 3

[17] Highsoft. Stacked column demo. https://www.highcharts.com/demo/column-stacked, 2018. Accessed: 2022-12-31. 7

[18] A. Hurst and S. Kane. Making "making" accessible. In *Proceedings of the 12th International Conference on Interaction Design and Children*, IDC '13, pp. 635–638. Association for Computing Machinery, New York, NY, USA, 6 2013. Accessed: 2021-09-03. 3

[19] C. Jung, S. Mehta, A. Kulkarni, Y. Zhao, and Y.-S. Kim. Communicating visualizations without visuals: Investigation of visualization alternative text for people with visual impairments. *IEEE Transactions on Visualization and Computer Graphics*, 28(1):1095–1105, Jan. 2022. doi: 10.1109/tvcg.2021.3114846 2

[20] J. Kim, A. Srinivasan, N. W. Kim, and Y.-S. Kim. Exploring chart question answering for blind and low vision users. In *Proceedings of the 2023 CHI Conference on Human Factors in Computing Systems*. ACM, Apr. 2023. doi: 10.1145/3544548.3581532 2

[21] N. W. Kim, S. C. Joyner, A. Riegelhuth, and Y. Kim. Accessible visualization: Design space, opportunities, and challenges. *Computer Graphics Forum*, 40(3):173–188, June 2021. doi: 10.1111/cgf.14298 9

[22] C.-Y. Lin and Y.-M. Chang. Increase in physical activities in kindergarten children with cerebral palsy by employing MaKey–MaKey-based task systems. *Research in Developmental Disabilities*, 35(9):1963–1969, Sept. 2014. doi: 10.1016/j.ridd.2014.04.028 3

[23] P. Louridas. Design as bricolage: anthropology meets design thinking. *Design Studies*, 20(6):517–535, Nov. 1999. doi: 10.1016/s0142-694x(98)00044-1 9

[24] A. Lundgard, C. Lee, and A. Satyanarayan. Sociotechnical considerations for accessible visualization design. In *2019 IEEE Visualization Conference (VIS)*. IEEE, Oct. 2019. doi: 10.1109/visual.2019.8933762 2, 7, 8, 9

[25] A. Lundgard and A. Satyanarayan. Accessible visualization via natural language descriptions: A four-level model of semantic content. *IEEE Transactions on Visualization and Computer Graphics*, 28(1):1073–1083, Jan. 2022. doi: 10.1109/tvcg.2021.3114770 2, 7, 8, 9

[26] D. Ma'ayan, W. Ni, K. Ye, C. Kulkarni, and J. Sunshine. How domain experts create conceptual diagrams and implications for tool design. In *Proceedings of the 2020 CHI Conference on Human Factors in Computing Systems*, CHI '20, p. 1–14. Association for Computing Machinery, New York, NY, USA, 2020. doi: 10.1145/3313831.3376253 3, 9

[27] K. Marriott, B. Lee, M. Butler, E. Cutrell, K. Ellis, C. Goncu, M. Hearst, K. McCoy, and D. A. Szafir. Inclusive data visualization for people with disabilities. *Interactions*, 28(3):47–51, Apr. 2021. doi: 10.1145/3457875 2, 3, 9

[28] B. E. Reid. The curb-cut effect and the perils of accessibility without disability. *SSRN Electronic Journal*, 2022. doi: 10.2139/ssrn.4262991 9

[29] Y. Rogers, J. Paay, M. Brereton, K. L. Vaisutis, G. Marsden, and F. Vetere. Never too old. In *Proceedings of the SIGCHI Conference on Human Factors in Computing Systems*. ACM, Apr. 2014. doi: 10.1145/2556288.2557184 3, 5

[30] Z. Sarsenbayeva, N. V. Berkel, E. Velloso, J. Goncalves, and V. Kostakos. Methodological standards in accessibility research on motor impairments: A survey. *ACM Computing Surveys*, 55(7):1–35, Dec. 2022. doi: 10.1145/3543509 3

[31] A. Satyanarayan, D. Moritz, K. Wongsuphasawat, and J. Heer. Vega-lite: A grammar of interactive graphics. *IEEE Transactions on Visualization and Computer Graphics*, 23(1):341–350, Jan. 2017. doi: 10.1109/tvcg.2016.2599030 3, 6, 8

[32] A. Sharif, S. S. Chintalapati, J. O. Wobbrock, and K. Reinecke. Understanding screen-reader users' experiences with online data visualizations. In *The 23rd International ACM SIGACCESS Conference on Computers and Accessibility*. ACM, Oct. 2021. doi: 10.1145/3441852.3471202 2

[33] A. Sharif and B. Forouraghi. evoGraphs — a jQuery plugin to create web accessible graphs. In *2018 15th IEEE Annual Consumer Communications Networking Conference CCNC*. IEEE, Jan. 2018. doi: 10.1109/ccnc.2018.8319239 2

[34] A. Sharif, O. H. Wang, A. T. Muongchan, K. Reinecke, and J. O. Wobbrock. VoxLens: Making online data visualizations accessible with an interactive JavaScript plug-in. In *CHI Conference on Human Factors in Computing Systems*. ACM, Apr. 2022. doi: 10.1145/3491102.3517431 2

[35] B. Shneiderman. The eyes have it: A task by data type taxonomy for information visualizations. In *The Craft of Information Visualization*, pp. 364–371. Elsevier, 2003. doi: 10.1016/b978-155860915-0/50046-9 7

[36] A.-I. Siean and R.-D. Vatavu. Wearable interactions for users with motor impairments: Systematic review, inventory, and research implications. In *The 23rd International ACM SIGACCESS Conference on Computers and Accessibility*. ACM, Oct. 2021. doi: 10.1145/3441852.3471212 3

[37] V. Sorge. Polyfilling accessible chemistry diagrams. In *Lecture Notes in Computer Science*, pp. 43–50. Springer International Publishing, 2016. doi: 10.1007/978-3-319-41264-1_6 2, 3

[38] J. R. Thompson, J. J. Martinez, A. Sarikaya, E. Cutrell, and B. Lee. Chart reader: Accessible visualization experiences designed with screen reader users. In *Proceedings of the 2023 CHI Conference on Human Factors in Computing Systems*. ACM, Apr. 2023. doi: 10.1145/3544548.3581186 2, 4, 8
[39] Visa. Visa Chart Components. https://github.com/visa/visa-chart-components, 2022. Accessed: 2022-12-01. 2
[40] WAI. Understanding success criterion 2.4.7: focus-visible. WCAG standard, W3C, 2016. Accessed: 2022-12-11. 6
[41] WAI. Understanding success criterion 4.1.2: name, role, value. WCAG standard, W3C, 2016. Accessed: 2022-03-04. 2
[42] WAI. Accessible rich internet applications (WAI-ARIA 1.1). Technical report, W3C, 2017. Accessed: 2022-12-11. 2
[43] WAI. Understanding success criterion 2.1.1: keyboard. WCAG standard, W3C, 2017. Accessed: 2022-12-11. 5
[44] WebAIM. WAVE, the web accessibility evaluation tool. https://wave.webaim.org/. Accessed: 2023-01-10. 7
[45] J. O. Wobbrock, S. K. Kane, K. Z. Gajos, S. Harada, and J. Froehlich. Ability-based design. *ACM Transactions on Accessible Computing*, 3(3):1–27, Apr. 2011. doi: 10.1145/1952383.1952384 9
[46] K. Ye, W. Ni, M. Krieger, D. Ma'ayan, J. Wise, J. Aldrich, J. Sunshine, and K. Crane. Penrose. *ACM Transactions on Graphics*, 39(4), Aug. 2020. doi: 10.1145/3386569.3392375 8
[47] J. Zong, C. Lee, A. Lundgard, J. Jang, D. Hajas, and A. Satyanarayan. Rich screen reader experiences for accessible data visualization. *Computer Graphics Forum*, 41(3):15–27, June 2022. doi: 10.1111/cgf.14519 2, 4, 6, 7, 8, 9